\documentclass[aps, prd, amsmath, floats, floatfix, twocolumn, superscriptaddress, nofootinbib, showpacs]{revtex4}

\usepackage{amssymb}
\usepackage{amsmath}
\usepackage{verbatim}
\usepackage{mathrsfs}
\usepackage{amsfonts}
\usepackage{latexsym}
\usepackage{epsfig}
\usepackage{epstopdf}
\usepackage{color}
\usepackage{graphicx,subfigure}
\usepackage{units}
\usepackage{hyperref}

\begin{document}


\definecolor{orange}{rgb}{0.9,0.45,0} 

\newcommand{\jmanuel}[1]{\textcolor{blue}{{\bf Jose M: #1}}}
\newcommand{\dario}[1]{\textcolor{red}{{\bf Dario: #1}}}
\newcommand{\alberto}[1]{\textcolor{green}{{\bf Alberto: #1}}}
\newcommand{\miguel}[1]{\textcolor{red}{{\bf Miguel: #1}}}

\renewcommand{\t}{\times}

\long\def\symbolfootnote[#1]#2{\begingroup%
\def\thefootnote{\fnsymbol{footnote}}\footnote[#1]{#2}\endgroup}


\newcommand{\tg}{\tilde{\gamma}}
\newcommand{\tG}{\tilde{\Gamma}}
\newcommand{\tA}{\tilde{A}}
\newcommand{\tR}{\tilde{R}}
\newcommand{\tnabla}{\tilde{\nabla}}

\newcommand{\hg}{\hat{\gamma}}
\newcommand{\hG}{\hat{\Gamma}}
\newcommand{\hA}{\hat{A}}
\newcommand{\hR}{\hat{R}}
\newcommand{\hD}{\hat{\Delta}}
\newcommand{\hnabla}{\hat{\nabla}}

\newcommand{\fg}{\mathring{\gamma}}
\newcommand{\fG}{\mathring{\Gamma}}
\newcommand{\fR}{\mathring{R}}
\newcommand{\fnabla}{\mathring{\nabla}}

\newcommand{\lb}{\pounds_{\vec{\beta}}}

\newcommand{\Lie}[2]{\pounds_{\vec{#1}}{#2}}
\newcommand{\p}{\partial}

\newcommand{\gpar}{{\hat{\gamma}}_\parallel}
\newcommand{\gper}{{\hat{\gamma}}_\perp}


\title{Cosmological nonlinear structure formation in full general relativity} 

\author{Jos\'e M. Torres} 
\affiliation{Instituto de Ciencias Nucleares, Universidad Nacional
  Aut\'onoma de M\'exico, Circuito Exterior C.U., A.P. 70-543,
  M\'exico D.F. 04510, M\'exico}

\author{Miguel Alcubierre}
\affiliation{Instituto de Ciencias Nucleares, Universidad Nacional
  Aut\'onoma de M\'exico, Circuito Exterior C.U., A.P. 70-543,
  M\'exico D.F. 04510, M\'exico}

\author{Alberto Diez-Tejedor} 
\affiliation{Santa Cruz Institute for Particle Physics and Department
  of Physics, University of California, Santa Cruz, CA, 95064, USA}  

\author{Dar\'{\i}o N\'u\~nez}
\affiliation{Instituto de Ciencias Nucleares, Universidad Nacional
  Aut\'onoma de M\'exico, Circuito Exterior C.U., A.P. 70-543,
  M\'exico D.F. 04510, M\'exico}


\date{\today}


\begin{abstract}
We perform numerical evolutions of cosmological scenarios using a
standard general relativistic code in spherical symmetry.  We
concentrate on two different situations: initial matter distributions
that are homogeneous and isotropic, and perturbations to those that
respect the spherical symmetry. As matter models we consider the case
of a pressureless perfect fluid, {\em i.e.} dust, and the case of a real massive
scalar field oscillating around the minimum of the potential. Both
types of matter have been considered as possible dark matter
candidates in the cosmology literature, dust being closely related to
the standard cold dark matter paradigm.  We confirm that in the linear
regime the perturbations associated with these types of matter grow in
essentially the same way, the main difference being that in the case
of a scalar field the dynamics introduce a cutoff in the power
spectrum of the density perturbations at scales comparable with the
Compton wavelength of the field. We also follow the evolutions well
beyond the linear regime showing that both models are able to form
structure. In particular we find that, once in the nonlinear regime,
perturbations collapse faster in a universe dominated by dust.
This is expected to delay the formation of the first structures in the
scalar field dark matter scenario with respect to the standard
cold dark matter one.
\end{abstract}




\maketitle


\section{Introduction}
\label{sec:introduction}

Structure formation as a consequence of the dynamical instability of
self-gravitating matter constitutes a cornerstone of modern physical
cosmology. As long as the deviations with respect to a homogeneous and
isotropic universe are small, it is always possible to develop a
self-consistent linear order theory where approximate analytic
solutions can be found~\cite{Peacock1998, Dodelson2003, Mukhanov2005}.
However, when the inhomogeneities grow and reach the nonlinear regime
those solutions fail and numerical techniques become necessary.

For the case of a universe dominated by classical particles the
evolution of the cosmological perturbations under the influence of
(Newtonian) gravity can be followed using N-body simulations (see {\em
  e.g.}~\cite{2012PDU.....1...50K} and references therein).  However,
when a description in terms of particles is not viable, we need other
techniques in order to address nonlinear structure formation. That is
the case, for instance, for models that consider dark matter and/or
dark energy as coherent macroscopic excitations of a boson field ({\em
  e.g.} nonthermal axionlike candidates for dark
matter~\cite{Dine:1982ah, Lee:1995af, Hu:2000ke, Sikivie:2009qn,
  2013arXiv1302.0903S}, quintessence models of dark
energy~\cite{Peebles:2002gy, Tsujikawa:2013fta},
$k$-essence~\cite{ArmendarizPicon:2000ah}), or models that propose a
modification of general relativity ({\em e.g.} MOND~\cite{Milgrom:2001ny,
  Bekenstein:2004ne}, Einstein-aether~\cite{Eling:2004dk,
  2008arXiv0801.1547J}, massive gravity~\cite{Hinterbichler:2011tt,
  deRham:2014zqa}), just to name some examples.

In this paper we adapt a numerical relativity code in spherical
symmetry to evolve cosmological scenarios and deal with the problem of
nonlinear structure formation in situations where a description of
matter in terms of fields is necessary. We use the OllinSphere2 code
described in Ref.~\cite{Alcubierre:2010is}, which solves the full Einstein
field equations in spherical symmetry coupled to different types of
matter. The code uses the standard 3+1 formulation of general
relativity with a strongly hyperbolic, well-posed system of evolution
equations~\cite{Alcubierre08a}.  Furthermore, it is regular in the
sense that the variables behave smoothly at the
origin~\cite{Ruiz:2007rs,Alcubierre:2010is}.  The OllinSphere2 code
has been used previously to address problems involving spherically
symmetric, self-gravitating compact objects in general
relativity~\cite{Alcubierre:2010ea,Ruiz:2012jt,Torres:2014fga}.

In this paper we use our code for the first time to study the
cosmological evolution of inhomogeneous distributions of matter, also
in spherical symmetry.  Particular attention is given to the cases of
a perfect fluid with no pressure (dust) and a massive scalar field
oscillating around the minimum of the potential showing that, in the
context of an expanding universe, small perturbations in both cases
can develop structures around otherwise homogeneous and isotropic
initial data.  This is indeed an interesting result: it has been
previously argued that, since the small perturbations of a canonical
scalar field propagate at the speed of light~\cite{Garriga:1999vw,
  Scherrer:2004au}, the coherent macroscopic excitations of a boson
field cannot give rise to a successful cosmological structure
formation.  This work, along with~\cite{Alcubierrepreparation}, aims
to provide evidence supporting the idea that a free classical massive
scalar field in a cosmological context can lead to large scale
structure formation.

This paper is organized as follows. In Section~\ref{sec:IVP} we review
the 3+1 formulation of general relativity. We also present the
Baumgarte-Shapito-Shibata-Nakamura (BSSN) system of evolution
equations implemented in the numerical code for the particular case of
spherical
symmetry~\cite{Shibata95,Baumgarte:1998te,Brown:2009dd,Alcubierre:2010is}.
Next, in Section~\ref{sec:num.evolutions} we consider spatially flat, homogeneous and
isotropic distributions of matter.  In
Section~\ref{sec:num.evolutions.II} we study the evolution of
spherically symmetric distributions of matter that represent small
perturbations away from homogeneity.  The results of our numerical
simulations are presented in Sections~\ref{sec.perfect fluid}
and~\ref{sec.scalar.field}, where we consider cosmologies dominated by
a pressureless perfect fluid, and by the coherent oscillations of a
real massive scalar field, respectively.  We find that the evolution
of the linear order perturbations in a universe dominated by a scalar
field is initially very similar to the case of a universe dominated by
dust, but with a cutoff in the power spectrum at the Compton
wavelength of the scalar particle (as was already known from previous
analytical results, {\it e.g.}~\cite{Ratra:1988bz, Hwang:1996xd, Matos:2000ng, Matos:2000ss, 
  Marsh:2010wq}). Once in the nonlinear regime, however, the evolution
of the inhomogeneities for both types of matter differs: while the
perturbations in the case of dust collapse very quickly, the growth of
the perturbations in the case of a scalar field is more gradual,
delaying the formation of the first structures in the universe.

\hspace{3mm}

We should mention at this point that during the final stage of
preparation of this paper we became aware of two recent studies that
address similar topics. On the one hand, the authors
of~\cite{2014NatPh..10..496S} (see also
Ref.~\cite{2014arXiv1407.7762S}) consider the problem of structure
formation in a universe dominated by an ultralight scalar field using
a hydrodynamic code in full 3D, but they assume from the beginning a
nonrelativistic, weak field approximation (see {\em e.g.}
Refs.~\cite{Adamek:2013wja, Bruni:2013mua, 2014arXiv1405.7006B,
  2014arXiv1409.6549R} for possible relativistic effects even in
N-body simulations, and Ref.~\cite{Urena-Lopez:2013naa} for
particular attention to the case of a scalar field). On the other hand
in Ref.~\cite{2014arXiv1409.3476R} the authors present a fully
relativistic numerical code for the study of cosmological problems in
spherical symmetry and use it to reproduce the analytical
Lemaitre-Tolman-Bondi solution. We believe the results presented in
this paper complement those previous studies.  See also
Refs.~\cite{Wainwright:2013lea, Wainwright:2014pta} for other
interesting recent applications of numerical relativity to cosmology,
now in the context of the early universe.


\section{Spacetime dynamics as an Initial Value Problem}
\label{sec:IVP}

In order to analyze the nonlinear behavior of the perturbations of a
Friedmann universe we find it convenient to formulate the evolution of
the cosmological spacetime as an initial value problem. A usual way of
doing that is to use the standard 3+1 formulation due to Arnowitt,
Desser and Misner (ADM)~\cite{Arnowitt62,York79}, in which the
spacetime is foliated into spatial hypersurfaces of constant time, and
the spacetime line-element is given by (throughout the paper we use a
signature $(-,+,+,+)$, and work in units such that
\mbox{$c=G=1$}):
\begin{equation}\label{eq:ADM}
 ds^2= \left( -\alpha^2 + \beta_i\beta^i \right) \: dt^2 + 2 \beta_i dt
 dx^i + \gamma_{ij} dx^i dx^j \, .
\end{equation}
Here $\alpha$ is a lapse function, $\beta^i$ the shift vector, and
$\gamma_{ij}$ the metric tensor of the three-dimensional spatial
hypersurfaces. In general all of them are functions of the spacetime
coordinates $(t,x^i)$. Spatial indexes are raised and lowered with the
spatial metric, {\em e.g.} $\beta_i = \gamma_{ij} \beta^j$, with
$\gamma^{ij} \gamma_{jk} \equiv \delta^i_k$.

In the 3+1 formalism the basic dynamical quantities are the metric of
the spatial hypersurfaces $\gamma_{ij}$, and the extrinsic curvature
tensor of those hypersurfaces $K_{ij}$ (see
{\em e.g.} Eq.~(\ref{eq:gammadot}) below for the definition of this
tensor). There are also four gauge functions, the lapse $\alpha$ and
the shift vector $\beta^i$, which are associated with the freedom in
the choice of the coordinate system. In terms of these variables the
Einstein field equations take the form of a constrained evolution
system similar to Maxwell's in the case of electromagnetism.

The constraints are obtained when projecting the Einstein field
equations along the normal direction to the spatial hypersurfaces. The
resulting expressions contain no time derivatives and are called the
Hamiltonian and momentum constraints:
\begin{subequations}\label{eq.constraints}
\begin{eqnarray}
{\cal H} &:=& R + K^2 - K_{ij} K^{ij}
- 16 \pi \rho = 0 \, , \label{eq:hamiltonian} \\
M^i &:=& \nabla_j \left( K^{ij} - \gamma^{ij} K \right)
- 8 \pi j^i = 0 \, . \label{eq:momentum}
\end{eqnarray}
\end{subequations}
In the above equations $R := \gamma^{ij} R_{ij}$ is the trace of the
Ricci tensor, \mbox{$K := \gamma^{ij} K_{ij}$} is the trace of the
extrinsic curvature, and $\nabla_i$ is the covariant derivative, all
of them defined with respect to the spatial metric $\gamma_{ij}$. As
usual $\rho$ and $j^i$ are the energy and momentum densities of matter
as measured by observers that move along the normal trajectories to
the spatial hypersurfaces, the so-called ``Eulerian observers'',
\begin{subequations}\label{source.grav.1}
\begin{eqnarray}
\rho &:=& n^\mu n^\nu T_{\mu \nu} \, , \label{eq:eulerian.density} \\
j^i &:=& - P^{i \mu} n^{\nu} T_{\mu \nu} \, .
\end{eqnarray}
\end{subequations}
The above quantities are defined in terms of the stress-energy tensor
of the matter fields, $T_{\mu \nu}$, with
\mbox{$n^{\mu}=(1/\alpha,-\beta^i/\alpha)$} the unit normal vector to
the spatial hypersurfaces of constant time, and $P_{\mu}^{\nu}:=
\delta_{\mu}^{\nu}+n_{\mu}n^{\nu}$ the projection operator.

The dynamics of the gravitational field is obtained when one projects
the Einstein equations onto the spatial hypersurfaces. Together with
the definition of the extrinsic curvature, the evolution equations are
given by
\begin{subequations}\label{eq.dynamical1}
\begin{eqnarray}
\partial_t \gamma_{ij} - \lb \gamma_{ij} &=& - 2 \alpha K_{ij} \, ,
\label{eq:gammadot} \\
\partial_t K_{ij} - \lb K_{ij} &=& - \nabla_i \nabla_j \alpha \nonumber \\
&+& \alpha ( R_{ij} + K K_{ij} - 2 K_{ik} K^k_j )
\nonumber \\
&+& 4 \pi \alpha \left[ \gamma_{ij} \left( S - \rho \right)
- 2 S_{ij} \right] \, .
\label{eq:Kdot}
\end{eqnarray}
\end{subequations}
Here $\lb$ is the Lie derivative along the direction of the shift
vector, and $S_{ij}$ the stress tensor measured by the Eulerian
observers,
\begin{equation}\label{source.grav.2}
S_{ij} := P^\alpha_i P^\beta_j T_{\alpha \beta} \; ,
\end{equation}
with $S := \gamma^{ij} S_{ij}$ its
trace. Equations~(\ref{eq.dynamical1}) must be completed with the
evolution equations for the matter fields and gauge functions, as well
as an appropriate set of initial data compatible with the
constraints~(\ref{eq.constraints}) and boundary conditions (see below
for details).

Although in principle one can solve the above system of equations and
obtain the future evolution of the spacetime, it turns out that the
mathematical properties of the system make it numerically unstable in
practice. A great amount of work in the field of numerical relativity
was focused on reformulating the system of evolution equations, by
introducing auxiliary fields and/or modifying the evolution equations
by adding terms proportional to the constraints, in order to render
the system stable. Here we will employ the BSSN
formulation~\cite{Shibata95,Baumgarte:1998te} adapted to curvilinear
coordinate systems~\cite{Alcubierre:2010is,Brown:2009dd}, that results
in a well-posed, strongly hyperbolic formulation of the 3+1 evolution
equations.  The BSSN formulation has proven to be very robust in
practice and is widely used in numerical relativity.

Next we review the main ingredients of the BSSN formulation (with
particular interest in the spherically symmetric case), and refer the
reader to the literature on the
subject~\cite{Alcubierre08a,Alcubierre:2010is} for further details on
its derivation.  We begin by introducing a conformal decomposition of
the spatial metric tensor of the form
\begin{equation}\label{eq.conformal.dec}
\gamma_{ij} := e^{4\chi}\hg_{ij} \, , \quad
\chi \equiv \ln\psi := \frac{1}{12} \ln(\gamma / \hg) \,.
\end{equation}
The conformal factor $e^{4\chi}$ introduced in Eq.~(\ref{eq.conformal.dec})
relates the physical volume element to the conformal one, which is
assumed to take initially its flat-space value.  In addition we have
defined the determinant of the two spatial metric tensors, the
physical $\gamma:={\rm det}(\gamma_{ij})$, and the conformal one,
$\hat{\gamma}:={\rm det}(\hat{\gamma}_{ij})$.

We will also introduce the traceless conformal extrinsic curvature
tensor,
\begin{equation}
\hA_{ij} := e^{-4\chi}\left( K_{ij} - \frac{1}{3}\gamma_{ij}K \right) \,,
\end{equation}
and a combination of contracted Christoffel symbols, which we will
call the ``connection vector'' from now on, of the form
\begin{equation}\label{eq.Delta}
\hat{\Delta}^i := \hat{\gamma}^{mn}(\hat{\Gamma}^{i}_{mn}
- \mathring{\Gamma}^{i}_{mn})\,.
\end{equation}
The over-ring in Eq.~(\ref{eq.Delta}) makes reference to some
arbitrary background metric which, for simplicity, can be taken to be
flat.  Note that even though neither $\hat{\Gamma}^{i}_{mn}$ nor
$\mathring{\Gamma}^{i}_{mn}$ are true tensors, their difference is in
fact a properly defined tensor, so that the quantity $\hat{\Delta}^i$
introduced above is a true vector.  In the BSSN formulation the
connection vector is considered to be an independent dynamical
variable, and its definition is treated as an additional constraint.

For the case of spherical symmetry the equations above simplify
considerably. In this case it is always possible to find a coordinate
system where the metric tensor of the spatial hypersurfaces takes the
form
\begin{equation}
\gamma_{ij} dx^i dx^j = e^{4 \chi} \left( \gpar dr^2
+ \gper r^2 d \Omega^2 \right)\, .
\label{eq:spheremetric}
\end{equation}
With this choice the connection vector has only the radial component
different form zero, \mbox{$\hat{\Delta}^i=(\hat{\Delta}^r,0,0)$}. The
traceless condition for the tensor $\hat{A}_{ij}$, together with the
assumption of spherical symmetry, now guarantees that
\mbox{$\hat{A}^r_r = - 2 \hat{A}^\theta_\theta = - 2
  \hat{A}^\varphi_\varphi := \hat{A}_\parallel$}. Note that the
dynamical quantities $\chi$, $\gpar$, $\gper$, $\hat{\Delta}^r$, and
$\hat{A}_{\|}$, as well as the trace of the extrinsic curvature $K$,
are functions only of time and the radial coordinate $r$.

Although the numerical code does not impose any particular gauge, from
now on and for the purposes of this paper it will be convenient to fix
the shift vector to zero.  In terms of the new variables the
Hamiltonian and momentum constraints take the form
\begin{subequations}\label{eq.constraint2}
\begin{eqnarray}
  \mathcal{H} &:=& R - \frac{3}{2} \hat{A}_\parallel^2 + \frac{2}{3} K^2 
  - 16 \pi \rho = 0 \, , \qquad
  \label{eq:sphere-ham} \\
  M_r &:=& \partial_r \hat{A}_\parallel - \frac{2}{3} \partial_r K
  + 6 \hat{A}_\parallel \partial_r \chi \nonumber{}\\
  & &+ \frac{3}{2} \hat{A}_\parallel \left( \frac{2}{r}
    + \frac{\partial_r \gper}{\gper} \right)
  - 8 \pi j_r = 0 \, ,
\label{eq:sphere-mom}
\end{eqnarray}
\end{subequations}
whereas for the evolution equations we find
\begin{subequations}\label{eq.dynamical2}
\begin{eqnarray}
\partial_t \chi &=& 
- \frac{1}{6} \alpha K \,, \\
\label{eq:sphere-chidot}
\partial_t \gpar &=& 
- 2 \alpha \gpar \hat{A}_\parallel \,, 
\label{eq:sphere-adot} \\
\partial_t \gper &=& 
\alpha \gper \hat{A}_\| \,, 
\label{eq:sphere-bdot}\\
\partial_t K &=& 
- \nabla^2 \alpha + \frac{1}{6}\alpha (9\hat{A}_\parallel^2 + 2 K^2) \nonumber{} \\ 
 &+& 4 \pi \alpha \left( \rho + S_\parallel + 2 S_\perp \right) \,, \quad
\label{eq:sphere-Kdot}\\
\partial_t \hat{A}_\parallel &=& 
- \left( \nabla^r \nabla_r \alpha - \frac{1}{3} \nabla^2 \alpha \right) 
+ \alpha \left( R_r^r - \frac{1}{3} R \right) \nonumber{} \\
&+& \alpha K \hat{A}_\parallel
- \frac{16}{3} \pi \alpha \left( S_\parallel - S_\perp \right) \, ,
\label{eq:sphere-Aadot}\\
\partial_t \hD^r &=& 
- \frac{2}{\hat{\gamma}_{\|}} \: \partial_r (\alpha \hat{A}_\parallel) 
+ 2 \alpha \hat{A}_\parallel \left( \hD^r
- \frac{3}{r\hat{\gamma}_{\perp}} \right) \nonumber{}\\
 &+& \frac{\alpha \xi}{\hat{\gamma}_{\|}} M_r \, .
\end{eqnarray}
\end{subequations}
In order to simplify the notation we have introduced the quantities
$S_\parallel := S^r_r$ and $S_\perp := S^\theta_\theta =
S^\varphi_\varphi$, with the momentum density given by
$j^i=(j^r,0,0)$. In the last equation above $\xi$ is an arbitrary
constant which must be such that $\xi > 1/2$ in order for the
evolution system to be strongly hyperbolic.  An optimal choice is
$\xi=2$, which eliminates the divergence of $\hat{A}_\parallel$ from
the evolution equation for $\hD^r$, and can also be shown to imply
that all propagating modes not directly related to the gauge choice do
so at the coordinate speed of light. This is the value that we use
in all the simulations below.~\footnote{In spherical symmetry there
  are no gravitational waves, so all propagating modes are either pure
  gauge modes, or modes associated with the constraints, neither of
  which represents the propagation of physical information.  Still,
  having the modes associated with the constraints propagating at the
  coordinate speed of light is advantageous from the numerical point
  of view.}

We shall also recast the matter field equations as a first order in
time system that we will solve simultaneously with the geometric
equations above; see Sections~\ref{sec.perfect fluid}
and~\ref{sec.scalar.field} below for our particular cases. Special
care should be taken when solving hydrodynamic scenarios since they
inherently develop shocks. In that case it is convenient to recast the
equations in a conservative form and apply high resolution
shock-capturing schemes~\cite{Font94}.

The formulation summarized above is well suited to standard numerical
techniques. In the code OllinSphere2 we discretize the spatial
derivative operators up to fourth order using standard centered
differences, and evolve forward in time using a method of lines with a
fourth order Runge-Kutta integration scheme. The coordinate
singularity associated with the origin in spherical symmetry is dealt
with by staggering the origin and defining auxiliary dynamical
functions that allow us to impose adequate regularity
conditions~\cite{Alcubierre:2010is}.  To improve stability during the
simulations we also add Kreiss-Oliger numerical dissipation terms to
the evolution equations~\cite{Gustafsson95}.

Finally, in order to find suitable initial data that satisfy the
constraints we employ different techniques depending on the scenario
we are studying, as will be explained in the sections below.


\section{Cosmological evolutions I: The homogeneous and isotropic universe}
\label{sec:num.evolutions}

A first step in order to perform numerical evolutions of nontrivial
matter distributions is to be able to reproduce the dynamics of a
homogeneous and isotropic background using the same algorithms. This
approach has been used previously in three dimensions to compare
numerical homogeneous spacetimes with their corresponding analytical
solutions \cite{Vulcanov:2001yv}. It is somewhat trivial since in a
homogeneous scenario all the spatial derivatives vanish, and a
consistent numerical implementation then reduces to solving the same
set of ordinary differential equations at every grid point.
Nevertheless, once we drop the homogeneous assumption we will need to
solve the full system and our methods will be optimal.

At this point we need to specify our gauge choice for the foliation,
that is, our ``slicing condition'' (remember that we are assuming that
the shift vector vanishes).  The simplest choice would be to take a
constant lapse, $\alpha=1$, which corresponds to the standard
``synchronous gauge'' used in cosmology.  In this case the coordinates
are comoving with the expansion of the universe, with the time
function $t$ representing the proper time measured by the comoving
observers.  One should be very cautious about this gauge choice since
it is well-known that in more general situations (black hole mergers,
gravitational collapse, compact objects) this choice of lapse,
commonly known as ``geodesic slicing'' in the numerical relativity
literature, tends to focus coordinate lines, which in turn leads to
coordinate singularities that quickly spoil the simulations (see %
{\em e.g.} Ref.~\cite{Alcubierre08a} for details). 
Furthermore, this choice 
formally spoils the strong hyperbolicity properties
of the evolution system, so it is undesirable for
studying the local dynamics of the gravitational field.

A better alternative is to extend our analysis by considering a more
general foliation of the Bona-Masso family~\cite{Bona94b},
\begin{equation}
\label{eq:BonaMasso}
\p_t \alpha = - \alpha^2 f(\alpha) K \,, \quad f(\alpha)>0 \,.
\end{equation}
The properties of this foliation depend on the choice of the function
$f(\alpha)$: choosing $f=0$ reduces to the case of geodesic slicing,
{\em i.e.} the synchronous gauge; taking instead $f=1$ turns out to be
equivalent to asking for the time coordinate to be a harmonic function
of spacetime and is known as ``harmonic slicing''; finally, choosing
\mbox{$f=2/\alpha$} corresponds to the standard ``1+log'' foliation
commonly used in black hole evolutions.

Since for the configurations we will be considering in this paper the
extrinsic curvature acquires a negative background value away from the
perturbations (we will be interested in the study of expanding
universes), the lapse function given in Eq.~(\ref{eq:BonaMasso}) will
be monotonically growing in those regions, and we must control this
behavior in order for our simulations to remain stable. It turns out
that choosing $f(\alpha)$ to be a constant with $f < 1/3$ keeps the
background coordinate speed of light $v_{\rm c} = \alpha / \psi^2$
always less than unity, which allows us to fulfill the Courant
stability condition. In the limiting case $f=1/3$ one recovers the
usual ``conformal time slicing'' $\alpha = \psi^2$ for the homogeneous
evolutions (more generally, taking $f(\alpha)$ constant one finds
$\alpha = \psi^{6f}$). For the results presented in this work we will
limit our choice to $f\simeq 1/3$ in order to compare with known
results, but
for the following discussion a particular lapse is not assumed.
(For practical purposes our coordinate time $t$ may
be identified with the usual cosmological conformal time $\eta$, that is $t=\eta$, 
but with the Big Bang not occurring necessarily at $\eta=0$ as explained below.)

In order to proceed we need to find initial data representing
homogeneous and isotropic distributions of matter. In any gauge with
vanishing shift vector that corresponds to
\begin{subequations}
\label{eq.initial data.1}
\begin{eqnarray}
\rho (t=t_0,r) &=& \bar{\rho}_0 \,, \\
j_r (t=t_0,r) &=& 0 \,, \label{momentum.bar.cero}\\
S_{\|} (t=t_0,r) &=& \bar{S}_{\|0}\,, \\
S_{\perp} (t=t_0,r) &=& \bar{S}_{\perp 0} = \bar{S}_{\|0} \,,
\label{pressure.bar.cero}
\end{eqnarray}
\end{subequations}
with $\bar{\rho}_0$ and $\bar{S}_{\|0}$ constants (note that the
equations for the matter fields can introduce some relations between
these two quantities).  From now on the subscript zero will make
reference to values measured at $t=t_0$, whereas the over-bar denotes
background quantities. That distinction will be relevant later for the
description of nontrivial matter distributions.

The constraint equations~(\ref{eq.constraint2}) can be satisfied
identically at $t=t_0$ if we choose
\begin{equation}\label{eq.initial.hom}
 \bar{K}_0^2 = 24\pi \bar{\rho}_0 \,, \quad
\bar{R}_0 = \bar{\hat{A}}_{\| 0} = 0 \,.
\end{equation}
The initial metric components $\bar{\chi}_0$,
$\bar{\hat{\gamma}}_{\|0}$, $\bar{\hat{\gamma}}_{\perp 0}$, and
$\bar{\alpha}_0$ are constants but otherwise arbitrary.  With no loss
of generality we can fix $\bar{\chi}_0=0$ ({\em i.e.} $\bar{\psi}_0=1$), and
$\bar{\hat{\gamma}}_{\|0}=\bar{\hat{\gamma}}_{\perp
  0}=\bar{\alpha}_0=1$.  The condition $\bar{R}_0=0$ guarantees that
the spatial hypersurface at $t=t_0$ is flat, so that
$\bar{\hat{\Delta}}^r_0=0$.

Introducing the above initial data into the evolution
equations~(\ref{eq.dynamical2}) we find that during the whole
evolution we must have
\begin{equation}
\bar{\hat{\gamma}}_{\|} = \bar{\hat{\gamma}}_{\perp} = 1 \,, \quad
\bar{\hat{A}}_{\|}=\bar{\hat{\Delta}}^r = 0 \,.
\end{equation}
On the other hand the conformal factor, the trace of the extrinsic
curvature and the lapse function must satisfy the ordinary
differential equations
\begin{subequations}\label{eqs.ev.hom}
\begin{eqnarray}
\partial_t \bar{\chi} &=&- \frac{1}{6} \: \bar{\alpha} \bar{K} \, ,
\label{eq:sphere-chidotF} \\
\partial_t \bar{K} &=& \frac{1}{3} \: \bar{\alpha}\bar{K}^2
+ 4 \pi\bar{\alpha}\left( \bar{\rho} + \bar{S} \right) \, , \label{eq:sphere-KdotF} \\
\partial_t \bar{\alpha} &=& -\bar{\alpha}^2 f(\bar{\alpha})\bar{K} \, ,
\end{eqnarray}
together with the Hamiltonian constraint (the momentum constraint is
now trivial)
\begin{equation}
 \bar{K}^2 = 24\pi \bar{\rho} \,.
\end{equation}
\end{subequations}
Clearly $\bar{\chi}$, $\bar{K}$, and $\bar{\alpha}$ are functions of
time only. Note that $\bar{S}$ is the trace of the stress tensor,
which in the case of a perfect fluid is just three times the value of
the pressure measured by the observers at rest with respect to the
fluid, $\bar{S}=3\bar{p}$.

Of course Eqs.~(\ref{eqs.ev.hom}) are just the standard Friedmann
equations in disguise. In order to see that it is in fact more
convenient to look at them in terms of the comoving time, $\tau = \int
\bar{\alpha}dt$, where the evolution of the lapse function decouples
and the other three equations simplify to~\footnote{Another
interesting possibility would be to use conformal time, $\eta=\int
\bar{\alpha}/\bar{\psi}^2 dt$, as we will do later in
Section~\ref{sec:num.evolutions.II}.}
\begin{subequations}\label{eqs.ev.hom.2}
\begin{eqnarray}
\partial_\tau \bar{\chi} &=& - \frac{1}{6} \: \bar{K} \, , \label{eq:sphere-chidotF.2} \\
\partial_\tau \bar{K} &=& \frac{1}{3} \: \bar{K}^2
+ 4 \pi\left( \bar{\rho} + \bar{S} \right) \,, \label{eq:sphere-KdotF.2} \\
\bar{K}^2 &=& 24\pi \bar{\rho} \,.
\end{eqnarray}
\end{subequations}
  From the three equations above
we can easily identify the scale factor $a$ and Hubble parameter
$H:=\partial_\tau a/a$,
\begin{equation}\label{eq:a,H}
a = \bar{\psi}^2 \,, \quad
H = 2 \partial_\tau \bar{\chi} = - \bar{K}/3 \,.
\end{equation}
It should be noted that in order to have initial data representing an
expanding universe one must choose the trace of the extrinsic
curvature to have a negative initial value, $\bar{K}_0=-3H_0<0$.  

To close the system we still need to specify the evolution equations
for the matter fields, as we will do later for the case of a perfect
fluid, Section~\ref{sec.perfect fluid}, and a scalar field,
Section~\ref{sec.scalar.field}.


\section{Cosmological evolutions II: The perturbations}
\label{sec:num.evolutions.II}

If we leave the symmetries of the homogeneous and isotropic background
aside (but still preserving the invariance under rotations about the
origin), the initial data in Eqs.~(\ref{eq.initial data.1}) can be
generalized to
\begin{subequations}\label{eq.initial data.2}
\begin{eqnarray}
 \rho(t=t_0,r)&=&\rho_0(r) \,,\\
 j_r(t=t_0,r)&=&j_{r0}(r) \,, \\
 S_{\|}(t=t_0,r)&=& S_{\|0}(r) \,, \\
 S_{\perp}(t=t_0,r)&=& S_{\perp 0}(r) \,.
\end{eqnarray}
\end{subequations}
Without loss of generality, we will also consider that the momentum
density vanishes at $t=t_0$, {\em i.e.} $j^r_0(r)=0$. This particular choice
implies that the matter is initially following the expansion at the
initial slice.

In order to compare our numerical simulations with previous analytical
results (obtained to linear order in a perturbation series), we will
consider the evolution of initial data that are not exactly
homogeneous and isotropic, but are very close to it. That is, we
assume that for any physical quantity $f(t,r)$ evaluated at $t=t_0$ we
can always write
\begin{equation}\label{eq:def.pert}
f_0(r) = \bar{f}_0 + \delta f_0(r) \,, 
\end{equation}
with $|\delta f_0(r)| \ll |\bar{f}_0|$. In our simulations the
functions $\delta f_0(r)$ at $t=t_0$ have either compact support on a
finite region, or decay very rapidly for large $r$.  We can then
safely identify $\bar{f}_0$ with the value of the function $f_0(r)$ at
the numerical boundary.  We will also choose the boundary of the
computational domain to be sufficiently distant so that the
perturbations never reach it during the simulation.

If we choose an initial slice with constant expansion $K_0^2 = 24\pi
\bar{\rho}_0$, together with $\hat{A}_{\| 0}=0$, the momentum
constraint~(\ref{eq:sphere-mom}) is still trivially satisfied, while
the Hamiltonian constraint~(\ref{eq:sphere-ham}) takes the form
\begin{equation}
\label{eq:Ham_pertur}
R_0 - 16\pi (\delta\rho_0) = 0 \,.
\end{equation}
The perturbation in the energy density, $\delta\rho_0$, now implies
that the initial spatial hypersurface cannot be flat. The three
dimensional Ricci scalar takes the general form \mbox{$R=\hR/\psi^4 -
  8 \: (\hnabla^2\psi)/\psi^5$}.  We can choose the spatial metric to
be initially conformally flat, so that
\mbox{$\hat{\gamma}_{\|0}=\hat{\gamma}_{\perp0}=1$},
$\hat{\Delta}^r_0=0$, and $\hR_0=0$.  The Hamiltonian constraint then
reduces to the following elliptic equation for the conformal factor,
\begin{equation}
\label{eq:Ham_pert2}
\hnabla^2\psi_0 + 2 \pi (\delta\rho_0) \psi_0^5 = 0 \,.
\end{equation}
In spherical symmetry this is just a second-order ordinary
differential equation. There are two linearly independent solutions to
Eq. (\ref{eq:Ham_pert2}); however, regularity at the origin
and an asymptotic decay to the background value,
\begin{equation}
\label{eq:Ham_bc}
\p_r \psi_0(r=0) = 0 \,, \quad \psi_0(r\rightarrow \infty) = 1 \,,
\end{equation}
singles out the physical solution. Note that if the perturbation
vanishes everywhere, $\delta\rho_0(r)=0$, then $\psi_0 (r) = 1$, and
we recover the background expression of the conformal factor.

To evaluate the perturbation $\delta f(t,r)$ of a given dynamical
quantity at any time $t>t_0$ in the evolution we simply remove
from the field $f(r)$ its value at the numerical boundary, where the
effect of the inhomogeneity has not arrived yet. Of course we can do
this only for a finite amount of evolution time, but as we have argued
previously this will be sufficient for the purposes of this paper.

\vspace{5mm}

Some words about how to compare our simulations with the standard
cosmological treatment of linear perturbations are in order here.  In
cosmology the metric tensor of a flat Friedmann universe with small
perturbations is usually parametrized in the form~\footnote{See for
  instance Section~7.1 in Ref.~\cite{Mukhanov2005}. Note however that
  we use a different signature for the spacetime metric, and a
  coordinate system adapted to the spherical symmetry.}
\begin{eqnarray}\label{eq.nearlyRW.sph}
ds^2 &=& a^2 \left[- \left( 1 + 2 \Phi \right) d\eta^2
- 2\partial_i B \: dx^i d\eta \right. \nonumber \\
&+& \left( 1 - 2\Psi -2 \partial^2_r E \right) dr^2 \nonumber \\
&+& \left. \left( 1-2\Psi-(2/r) \: \partial_r E \right) r^2 d\Omega^2 \right] \, . 
\end{eqnarray}
We restrict our attention to the scalar sector of the perturbations:
vector and tensor modes cannot be excited in a spherically symmetric
configuration and for that reason they have not been included here.
Under these assumptions the fields $\Phi$, $B$, $\Psi$ and $E$ in
Eq.~(\ref{eq.nearlyRW.sph}) measure the small deviations of a
spacetime with respect to an exact flat homogeneous and isotropic
universe.

The spacetime length-element in the numerical code, on the contrary, 
can be expressed in the form
\begin{equation}\label{eq.metrica4}
 ds^2 = a^2 \left[-\frac{\psi^4}{a^2} d\eta^2
+ \frac{\psi^4}{a^2} \left(\hat{\gamma}_{\|}dr^2
+ \hat{\gamma}_{\perp}r^2d\Omega^2\right) \right] \,.
\end{equation}
Here the scale factor is identified from the background universe,
Eq.~(\ref{eq:a,H}) above, with $\eta=\int \bar{\alpha}/\bar{\psi}^2 dt$ the
cosmological time in conformal coordinates as a function of the time
coordinate $t$ of the code.  Comparing Eqs.~(\ref{eq.nearlyRW.sph})
and~(\ref{eq.metrica4}) we can easily identify $1 + 2
\Phi=e^{4\delta\chi}$, $B=0$, $1 - 2\Psi - 2\partial^2_r E =
e^{4\delta\chi} \hat{\gamma}_{\|}$, and $1-2\Psi-(2/r) \partial_r E =
e^{4\delta\chi}\hat{\gamma}_{\perp}$.  Working to first-order in the
perturbations, and after some algebra, we finally obtain
\begin{subequations}\label{eq:defs.pert.metric}
\begin{eqnarray}
 \Phi &=& 2\delta\chi\,, \\ 
 B &=& 0 \,, \\ 
 \Psi &=& -\int^r\frac{dr_1}{2r_1}\left(\delta\hat{\gamma}_{\perp} -
 \delta\hat{\gamma}_{\|}\right) -
 \frac{1}{2}\delta\hat{\gamma}_{\perp} - 2\delta\chi \,,  \label{eq:Psi}\\ 
 E &=& \int^r r_2
 dr_2 \int^{r_2}\frac{dr_1}{2r_1}\left(\delta\hat{\gamma}_{\perp} -
 \delta\hat{\gamma}_{\|}\right) \,. \label{eq:E}
\end{eqnarray}
\end{subequations}
The constants of integration in Eqs.~(\ref{eq:Psi}) and~(\ref{eq:E})
are fixed in order to have regular values of $\Psi$ and $E$ that match
the homogeneous background at the boundary.  Remember that we
previously fixed the shift vector to zero and for that reason the
field $B$ vanishes.

Note that the perturbations in the metric tensor as defined in
Eqs.~(\ref{eq.nearlyRW.sph}) and~(\ref{eq:defs.pert.metric}) depend on
the particular choice of coordinates.  However, it is always possible
to define a set of gauge-invariant functions of the form
\begin{subequations}\label{eq.gauge.metric}
 \begin{eqnarray}
  \Phi_{\textrm{g.i.}} &=& \Phi - (1/a) \left[ a(B-E') \right]' \,, \\
  \Psi_{\textrm{g.i.}} &=& \Psi + \mathcal{H}(B-E') \,,
 \end{eqnarray}
\end{subequations}
that do not change under general (first order) coordinate
transformations.  Here the prime denotes the derivative with respect
to the conformal time, and $\mathcal{H}:=a'/a$.~\footnote{In this
  paper we will consider matter fields that do not develop anisotropic
  stresses (to linear order); this guarantees that
  $\Phi_{\textrm{g.i.}}= \Psi_{\textrm{g.i.}}$ (to this same order
  in the perturbation series).  The field $\Phi_{\textrm{g.i.}}$ is
  analogous to the gravitational potential in the nonrelativistic
  theory and for that reason it is usually known as the Newtonian
  potential.}  A similar expression for a gauge-invariant perturbation
in the energy density takes the form
\begin{equation}\label{eq.gauge.density}
 \delta\rho_{\textrm{g.i.}} = \delta\rho -\bar{\rho}'(B-E')\,.
\end{equation}
Equations~(\ref{eq.gauge.metric}) and~(\ref{eq.gauge.density}) were
reported for the first time by Bardeen in Ref.~\cite{Bardeen:1980kt},
and since then they have been frequently used to characterize
cosmological linear order perturbations. They involve derivatives with
respect to the conformal time, so that then they are in fact
numerically nontrivial. However, they can be written explicitly in
terms of other fields by means of the evolution equations. In
particular, for the time derivatives of the perturbations in the
metric tensor we obtain, up to linear order,
\begin{subequations}
\begin{eqnarray}
\label{eq:gauge.Eprimes}
E^\prime &=& \frac{3}{2} \bar\psi^2 \int^r r_2dr_2 \int^{r_2} \frac{dr_1}{r_1} \hat A_\| \: ,
\label{eq:Eprime} \\
E^{\prime\prime} &=& \frac{3}{2} \frac{\bar\psi^4}{\bar\alpha} \int^r r_2 dr_2
\int^{r_2} \frac{dr_1}{r_1} \left(\partial_t \hat A_\|
- \frac{1}{3}\bar \alpha \bar K \hat A_\| \right) . \hspace{8mm} \label{eq:Epprime}
\end{eqnarray}
\end{subequations}
The explicit form of ${\bar\rho}^{\prime}$ can be found in a similar
way once the matter fields are specified.

At this point it is important to emphasize once again that the code
OllinSphere2 evolves initial data in terms of some dynamical fields
$f(t,r)$, and it does not distinguish between background values,
$\bar{f}(t)$, and their perturbations, $\delta f(t,r)$; see for
instance Eqs.~(\ref{eq.constraint2}) and~(\ref{eq.dynamical2}) above.
However, the definitions in this section are convenient in order to
compare the numerical evolutions presented here with previous
analytical results obtained when ``the perturbations are not to
large'', as we will do next in Section~\ref{sec.perfect fluid} for the
case of a pressureless perfect fluid.


\section{Pressureless perfect fluid (dust)}
\label{sec.perfect fluid}

At large scale dark matter is usually described in terms of a perfect
fluid with equation of state \mbox{$p=0$}, known generically as
``dust'' in relativity. The energy-momentum tensor of a perfect fluid
with no pressure takes the form $T_{\mu\nu} = \rho_{\textrm{rest}}
u_\mu u_\nu$, where $u_{\mu}$ is the four-velocity of the fluid
elements and $\rho_{\textrm{rest}}:=T_{\mu\nu}u^{\mu}u^{\nu}$ is the
energy density measured by the observers comoving with the fluid.
Note that, in general, the energy density measured by the observers at
rest with respect to the fluid does not coincide with the energy
density measured by the Eulerian observers,
Eq.~(\ref{eq:eulerian.density}).

For the case of a universe dominated by a perfect fluid with vanishing
pressure there is in fact an analytic expression for the scale factor
as a function of the conformal cosmological time,~\footnote{See
  {\em e.g.} Eqs.~(1.77), (7.53), and~(7.54) in Ref.~\cite{Mukhanov2005}.}
\begin{equation}\label{eq:scale.factor.dust}
a(\eta) = a_0 \left( \frac{\eta}{\eta_{\textrm{BB}}} - 1 \right)^2 \,,
\end{equation}
as well as for the gauge-invariant perturbations in the metric tensor,
\begin{subequations}\label{eq.sol.linear.p=0}
\begin{eqnarray}\label{eq.Psi.linear}
\Phi_{\textrm{g.i.}}(\eta,\vec{x}) &=& \Psi_{\textrm{g.i.}}(\eta,\vec{x}) \nonumber \\
&=& C_1(\vec{x}) + C_2(\vec{x}) \left( 
\frac{\eta}{\eta_{\textrm{BB}}} - 1 \right)^{-5} , \hspace{5mm}  
\end{eqnarray}
and the contrast in the energy density, 
\begin{eqnarray}\label{eq.contrast.linear}
\frac{\delta\rho_{\textrm{g.i.}}}{\bar{\rho}} \: (\eta,\vec{x})
= \frac{1}{6} \bigg\{ \left[ \Delta C_1 (\vec{x})
\left( \eta - \eta_{\textrm{BB}} \right)^2 - 12 C_1(\vec{x}) \right] \hspace{7mm} \nonumber \\
+ \left[\Delta C_2(\vec{x}) \left(\eta-\eta_{\textrm{BB}} \right)^2
+ 18 C_2(\vec{x}) \right] \hspace{-1mm}
\left( \frac{\eta}{\eta_{\textrm{BB}}} -1 \right)^{-5} \hspace{-1mm} \bigg\} . \hspace{10mm}
\end{eqnarray}
\end{subequations}
Here the origin of conformal time $\eta=0$ has been chosen
arbitrarily, with the the singularity at the Big Bang fixed to
\mbox{$\eta=\eta_{\textrm{BB}}=-[3/(2\pi a_0^2\bar{\rho}_0)]^{1/2}$}.
Also, $C_1(\vec{x})$ and $C_2(\vec{x})$ are arbitrary functions of
the spatial coordinates that depend on the initial perturbations and
can be determined from the initial data through
\begin{subequations}\label{eq.C1,C2}
\begin{eqnarray}
C_1(\vec x) &=& \Psi_{\textrm{g.i.}}(0,\vec x) - \frac{\eta_{\textrm{BB}}}{5}
\: \Psi_{\textrm{g.i.}}^\prime(0,\vec x) \,, \\
C_2(\vec x) &=& - \frac{\eta_{\textrm{BB}}}{5}  \: \Psi_{\textrm{g.i.}}^\prime(0,\vec x) \,. 
\end{eqnarray}
\end{subequations}

Notice that, after some transient time, the Newtonian potential
remains constant, $\Phi_{\textrm{g.i.}}=\Psi_{\textrm{g.i.}}\sim
C_1(\vec{x})$. The behavior of the contrast in the energy density, on
the other hand, depends on the size of the perturbations. Those modes
shorter than the Hubble radius, $k r_H \gg 1$, grow in time as
$(\eta-\eta_{\rm BB})^2$ (the square of the comoving Hubble radius
\mbox{$r_H\equiv1/\mathcal{H}=(\eta-\eta_{\rm BB})/2$}), whereas
Fourier modes that are initially larger than the Hubble radius,
\mbox{$k r_H(\eta=0) \ll 1$}, remain essentially frozen until the horizon
reaches them.

\vspace{5mm}

To deal with the numerical integration of the fluid-dynamical
equations forward in time, our code has an implementation of the
``Valencia formulation'' of general relativistic
hydrodynamics~\cite{Font94,Font:2008fka}, where the fluid dynamical
variables are given in terms of the so called ``conserved quantities''
\begin{subequations}
\begin{eqnarray}
\label{eq:ValenciaVars}
D\, &:=&  \rho_{\textrm{rest}} W \,, \\
\mathcal{S}^i &:=& \rho_{\textrm{rest}} h W^2 v^i \,, \\
\mathcal{E}\; &:=& \rho_{\textrm{rest}} h W^2 - p - D \,.
\end{eqnarray}
\end{subequations}
In these expressions $h := 1 + \epsilon + p/\rho_{\textrm{mass}}$ is
the specific enthalpy, with $\rho_{\textrm{mass}}$ the rest mass
density and $\epsilon$ the specific internal energy defined as
\mbox{$\epsilon := \rho_{\textrm{rest}}/\rho_{\textrm{mass}} - 1$}. We
have also introduced the variable $W := - u^{\mu} n_{\mu} = \alpha u^0
= 1/\sqrt{1 - \gamma_{ij} v^i v^j}$, which is just the Lorentz factor
associated with the velocity of the fluid elements $v^i$ measured by
the Eulerian observers.  The conserved quantities have the following
physical interpretation: $D$ is the rest mass density measured by the
Eulerian observers, $\mathcal{S}^i$ is the momentum density measured
by the same observers, while $\mathcal{E}$ is the difference between
the total energy density and the rest mass density measured in the
same Eulerian frame, $\mathcal{E}=\rho-D$.  We have defined the
conserved quantities for the general case of nonzero pressure;
however, notice that for a pressureless fluid we will in fact have
$p=0$ and $\rho_{\textrm{mass}}=\rho_{\textrm{rest}}$, so that
$\epsilon=0$ and $h=1$.

An important issue in the Valencia formulation of relativistic
hydrodynamics is the fact that the conversion from the primitive
variables $(\rho_{\textrm{rest}},\epsilon,v^i)$ to the conserved
quantities $(D,\mathcal{E},\mathcal{S}^i)$ is given in terms of
transcendental equations.  However, for the situation of interest here
this is not relevant since for dust we have $p=0$, $\epsilon=0$, and
$h=1$, so that
\begin{equation}
v^i = \mathcal{S}^i/(\mathcal{E} + D)  \, , \quad
\rho_{\textrm{rest}} = D/W \, . 
\end{equation}

In terms of the conserved quantities, the evolution equations for a
pressureless perfect fluid can be written in conservative form as
\begin{subequations}
\begin{eqnarray}
\label{eq:ValenciaEvol}
\p_t D\, - \Lie{\beta}{D} \, &=&-\nabla_k \left( \alpha D v^k \right)
+ \alpha K D \,, \\ 
\p_t \mathcal{S}^i - \Lie{\beta}{\mathcal{S}^i} &=&
- \nabla_k \left( \alpha \mathcal{S}^i v^k \right)
+ \alpha K \mathcal{S}^i \nonumber \\ 
&-& \left( \mathcal{E} + D \right) \nabla^i \alpha \,, \\
\p_t \mathcal{E}\, - \Lie{\beta} {\mathcal{E}} \, &=& -\nabla_k \left( \alpha \mathcal{E}v^k \right)
+ \alpha K \mathcal{E} \nonumber \\
&+& \left( \mathcal{E} + D \right) \left( \alpha v^m v^n K_{mn} - v^m \p_m \alpha \right) \,.
\hspace{8mm}
\end{eqnarray}
\end{subequations}
On the other hand, the sources of the gravitational field given in
Eqs.~(\ref{source.grav.1}) and~(\ref{source.grav.2}) take a
particularly simple form for a pressureless fluid,
\begin{subequations}
\begin{eqnarray}
\rho &=& \mathcal{E} + D \,, \\
j^i &=& \mathcal{S}^i \,, \\
S_{ij} &=& v_i \mathcal{S}_j \,.
\end{eqnarray}
\end{subequations}
Our numerical code integrates, in spherical symmetry, these equations
forward in time simultaneously with the integration of the spacetime
variables.  While the spacetime integration can be dealt with a simple
method of lines, the equations for the fluid are integrated with
advanced shock-capturing numerical methods (see {\em e.g.}~\cite{Leveque92}).

As already mentioned, we perform all our simulations with a vanishing
shift vector and a slicing condition of the Bona-Masso family,
Eq.~\eqref{eq:BonaMasso}. In particular, we chose \mbox{$f=0.3332$} in
order to be close to the case that reduces to the conformal time
slicing for a homogeneous distribution.
Taking this into account we can compare directly the result of our evolutions with
the analytical expressions such as those in Eqs.~\eqref{eq:scale.factor.dust}
and~\eqref{eq.sol.linear.p=0}.  We use a numerical grid of $10^4$
points, with a spacing of $\Delta r = 0.1$.

For the case of a perfect fluid with no pressure and vanishing initial
velocity with respect to the comoving coordinates,
$v^r(t=t_0)=0$, the initial data of Eqs.~(\ref{eq.initial
  data.2}) simplifies to
\begin{subequations}\label{eq.initial data.3}
\begin{eqnarray}
\rho(t=t_0,r) &=& \rho_0(r) \,,\\
j^r(t=t_0,r) &=& 0 \,, \\
S_{\|}(t=t_0,r) &=& 0 \,, \\
S_{\perp}(t=t_0,r) &=& 0 \,.
\end{eqnarray}
\end{subequations}
Notice that since initially $v^r=0$, the energy density measured by
the Eulerian observers coincides with the one measured by the
observers comoving with the fluid, so that
\mbox{$\rho_0(r)=\rho_{\textrm{rest}0}(r)=D_0(r)$}, $S^r_0 =
\mathcal{E}_0=0$.

Once the initial profile for the energy density $\rho_0(r)$ is given,
and it is decomposed into a background and a perturbation,
\begin{subequations}\label{eq.density.initial.gen}
\begin{equation}\label{eq.density.initial}
\rho_0(r) = \bar{\rho}_0 + \delta\rho_0(r) \,,
\end{equation}
all that one needs to do is to specify an initial geometry consistent
with such a matter distribution, as we learned previously in
Sections~\ref{sec:num.evolutions} and~\ref{sec:num.evolutions.II}.  In
order to construct initial data corresponding to a localized
inhomogeneity living in an expanding, otherwise homogeneous and
isotropic universe, we parametrize the initial perturbation of the
energy density as
\begin{equation}\label{par.pertur2}
\delta\rho_0(r) = \delta\rho_{*}
\left( 1- \frac{11 r^2}{3L^2}\right) \left[\left(1-\frac{r}{L}\right)
\left( 1 + \frac{r}{L} \right) \right]^3 \: , 
\end{equation}
\end{subequations}
for $0<r<L$, and $\delta\rho_0(r) =0$ for $r \geq L$.  Here
$\delta\rho_{*}$ and $L$ are an amplitude and a length-scale for the
initial perturbation, respectively (see for instance the first panels
in Figures~\ref{fig:Dust_L.jmcont} and~\ref{fig:Dust_L2.jmcont} below for 
particular realizations). We have chosen such a
parametrization in order to guarantee that the perturbation has both
compact support and integrates to zero, $\int_V \delta\rho_0(r) \:
d^3\vec{x}=0$. Our motivation here is that the perturbation should
correspond to a redistribution of matter within a flat universe, and
not to an addition of extra matter. As long as $\delta\rho_{*}\ll
\bar{\rho}_0$ we should have a cosmological evolution that can be well
described in terms of linear perturbation theory,
Eqs.~(\ref{eq.sol.linear.p=0}) above, at least for a finite period of
time until the nonlinear evolution becomes relevant.

\begin{figure}
  \centering
  \includegraphics[width=0.49\textwidth]{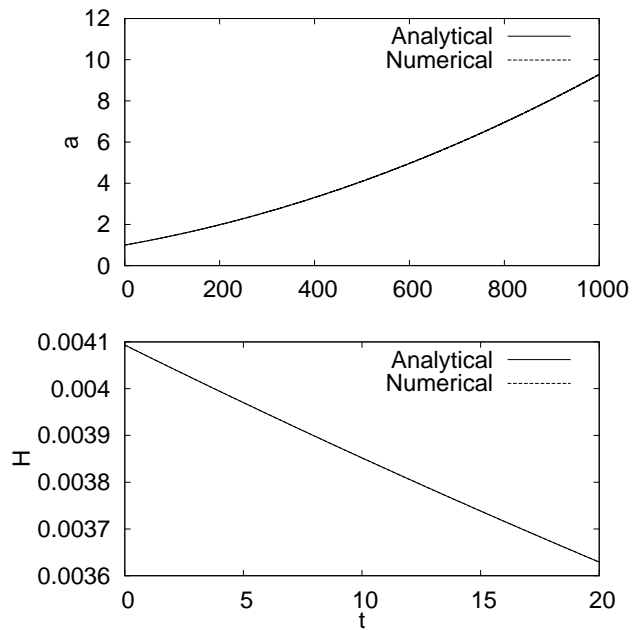}
  \caption{Evolution of the scale factor $a$ (top panel) and Hubble
    parameter $H$ (bottom panel), as functions of coordinate time $t$,
    for an expanding homogeneous and isotropic universe dominated by a
    perfect fluid with no pressure (dust). Here we have chosen
    $\rho_0=2\times10^{-6}$ and $\delta\rho_*=0$. We show both the
    numerical evolution (dotted-line), and the analytic solution of
    Eq.~(\ref{eq:scale.factor.dust}) (solid-line). The scale is zoomed
    in near the origin for the bottom panel in order to appreciate the
    detail and for comparison below.}
  \label{fig:Dust_a_evol}
\end{figure}

Figure~\ref{fig:Dust_a_evol} shows, for the case of a homogeneous and
isotropic universe with \mbox{$\bar{\rho}_0=2\times10^{-6}$} and
$\delta\rho_{*}=0$, 
a comparison of the evolution of the scale factor obtained with our
numerical implementation against the analytical solution for a
Friedmann universe given in Eq.~(\ref{eq:scale.factor.dust}). Since we
are interested in the case of an expanding configuration we need to
choose the negative root of the trace of the extrinsic curvature,
$K_0=-1.228\times 10^{-2}$.
As can be seen from the figure, the numerical
result is very accurate and converges consistently with the
discretization order of the integration method.~\footnote{Numerically we work with
  dimensionless coordinates and quantities. In order to recover the
  physical quantities we need an arbitrary length-scale $\lambda$ such
  that \mbox{$x^\mu_{\rm phys} = \lambda x^\mu$}; then the numerical
  solutions obtained can be related to the physical ones by dividing
  the derivatives of the extrinsic curvature $K_{ij}$ and connection vector $\hD^i$ by $\lambda$, and
  the matter sources $\rho$, $j^i$ and $S_{ij}$ by $\lambda^2$. For a
  pressureless fluid there is not an a priori choice for this scale,
  which is an advantageous fact that enables us to make comparisons
  with other models by choosing an appropriate physical scale.} 

We analyze now the evolution of the perturbed case. For this scenario
we have chosen an initial background density of
$\bar{\rho}_0=2.0\times 10^{-6}$, with an amplitude for the
perturbation of $\delta\rho_{*}=1.2\times 10^{-8}$. We have fixed the
characteristic length-scale in Eqs.~(\ref{eq.density.initial.gen}) to
$L=50$. The parameters have
been chosen in order to have a perturbation whose length-scale is well
inside the Hubble radius, with an initial contrast in the energy
density that allows us to compare the numerical evolutions with the
analytical results obtained in the linear regime.

\begin{figure}[t]
  \centering
  \includegraphics[width=0.49\textwidth]{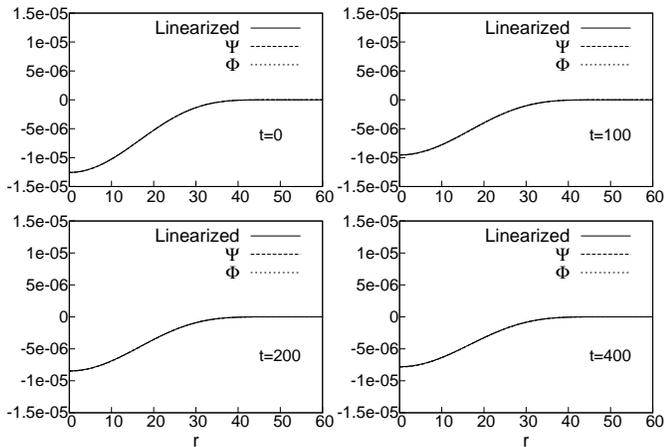}
  \caption{Snapshots of the evolution of the gauge-invariant
    perturbations in the metric tensor, $\Phi_{\textrm{g.i.}}$ and
    $\Psi_{\textrm{g.i.}}$, as functions of the comoving radial
    coordinate, $r$.  Here we have chosen initial data of the
    form in Eq.~(\ref{eq.density.initial.gen}) with $\bar{\rho}_0=2.0\times
    10^{-6}$, $\delta\rho_*=1.2\times10^{-8}$ and \mbox{$L=50$}. We
    show the results of the numerical evolution,
    $\Psi_{\textrm{g.i.}}$ (dotted-line), $\Phi_{\textrm{g.i.}}$
    (dashed-line), together with the analytic solution obtained in the
    linear regime, Eq.~(\ref{eq.Psi.linear}), (solid-line).}
  \label{fig:Dust_L.jmpot}
\end{figure}

In Figure~\ref{fig:Dust_L.jmpot} we show the numerical evolution of
the gauge-invariant metric perturbations, $\Phi_{\textrm{g.i.}}$ and
$\Psi_{\textrm{g.i.}}$, along with the evolution expected from the
linear order analysis, Eq.~(\ref{eq.Psi.linear}).  The functions
$C_1(\vec x)$ and $C_2(\vec x)$ in Eq.~(\ref{eq.Psi.linear}) have been
determined from the initial data using the expressions in
Eqs.~(\ref{eq.C1,C2}).  Throughout the whole evolution the relation
\mbox{$\Psi_{\rm g.i.}=\Phi_{\rm g.i.}$} is satisfied up to
numerical error, and converges consistently with the discretization
order.  Also, the numerical results coincide remarkably well with
those predicted by the analytic expressions obtained in the linear
regime. Note that after some time, in this case at $t\sim 300$, the
Newtonian potential approaches its asymptotic value $C_1(\vec x)$.

\begin{figure}[t]
  \centering
  \includegraphics[width=0.49\textwidth]{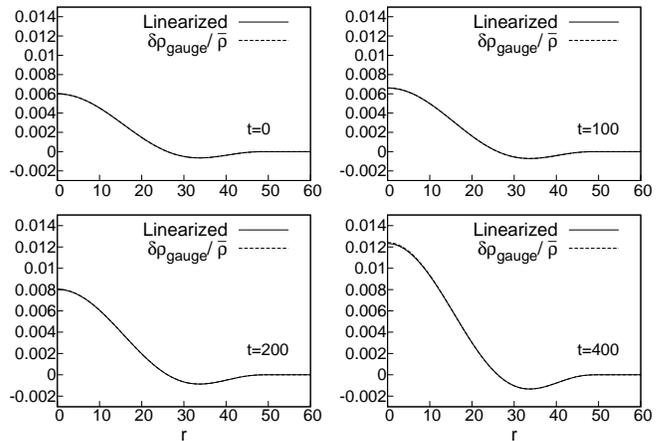}
  \caption{Snapshots of the contrast in the energy density,
    $\delta\rho_{\textrm{g.i.}}/\bar{\rho}$, for the same initial
    data as in Figure~\ref{fig:Dust_L.jmpot}. We show the results of
    the numerical evolution (dashed-line), together with the analytic
    solution obtained in the linear regime,
    Eq.~(\ref{eq.contrast.linear}), (solid-line).}
  \label{fig:Dust_L.jmcont}
\end{figure}

Figure~\ref{fig:Dust_L.jmcont} shows, for the same initial data as in
Figure~\ref{fig:Dust_L.jmpot}, the numerical evolution of the contrast
in the energy density, $\delta\rho_{\textrm{g.i.}}/\bar{\rho}$, and
we compare it with the evolution predicted with the linear order
theory, Eq.~(\ref{eq.contrast.linear}). Since the characteristic size
of the initial perturbation is well inside the Hubble radius, all the
significant modes contributing to the inhomogeneity grow at the same
rate and the profile is only rescaled as time moves forward.  As can
be appreciated from the figure, the evolution of the density contrast
also coincides very well with the prediction obtained from the
linearized theory.

\begin{figure}[t]
  \centering
  \includegraphics[width=0.49\textwidth]{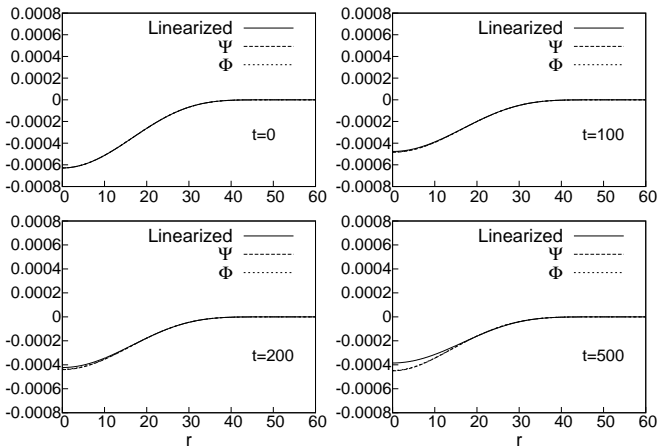}
  \caption{Similar to Figure~\ref{fig:Dust_L.jmpot} but now for
    initial data with $\bar{\rho}_0=2.0\times 10^{-6}$,
    $\delta\rho_*=1.2\times10^{-7}$ and \mbox{$L=50$} in
    Eq.~(\ref{eq.density.initial.gen}). Note that, after some
    transient time $t\sim 200$, deviations from the linearized
    predictions become apparent.}
  \label{fig:Dust_L2.jmpot}
\end{figure}

\begin{figure}[t]
  \centering
  \includegraphics[width=0.49\textwidth]{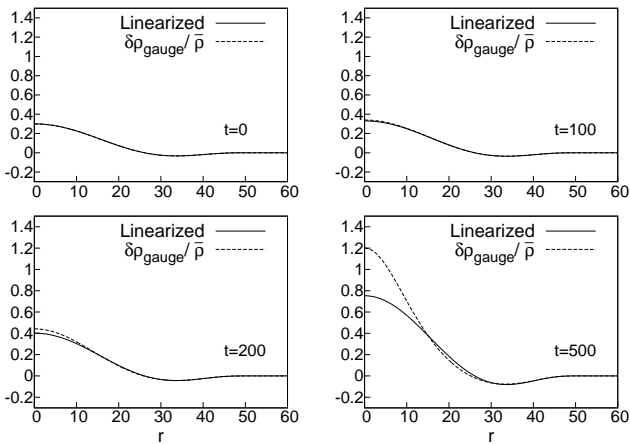}
  \caption{Similar to Figure~\ref{fig:Dust_L.jmcont} but now for the
    same initial data of Figure~\ref{fig:Dust_L2.jmpot}.  Note again
    that, after some transient time $t\sim 200$, deviations from
    the linearized predictions become apparent.}
  \label{fig:Dust_L2.jmcont}
\end{figure}

Our numerical results confirm what was already known from linear theory
for the behavior of small perturbations.  However, in a universe
dominated by dust the inhomogeneities are expected to grow
indefinitely as the universe expands.  This poses a problem for the
linearized approximation since the natural evolution will drive the
system away from this regime.  In order to dive into the nonlinear
regime we have performed simulations with the same background settings
as those in Figures~\ref{fig:Dust_L.jmpot} and~\ref{fig:Dust_L.jmcont}
above, but with a larger perturbation amplitude,
\mbox{$\delta\rho_{*}=6\times 10^{-7}$}, in order to have an initial
value for the energy density contrast of order $ 10^{-1}$. The new
simulations proceed initially in a similar way to the previous ones,
see Figures~\ref{fig:Dust_L2.jmpot} and~\ref{fig:Dust_L2.jmcont}.
However, as expected, we soon find an appreciable deviation from the
predictions of the linear theory. This deviation grows as time
proceeds, although it is interesting to note that the regions that
have not entered the nonlinear regime, $r\gtrsim 15$ in the figures,
are still described reasonably well by the linear order expressions.

\begin{figure}[t]
  \centering
  \includegraphics[width=0.45\textwidth]{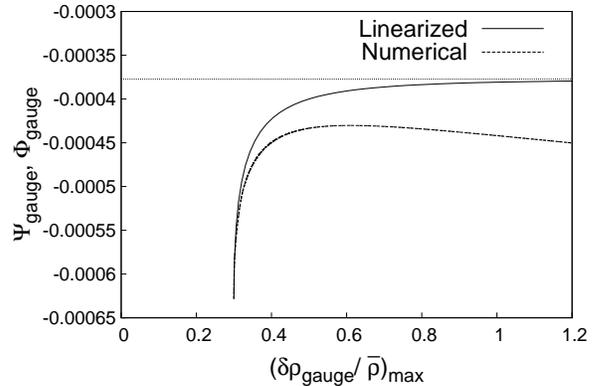}
  \caption{Central value of the potentials $\Psi_{\rm g.i.}$ and
    $\Phi_{\rm g.i.}$ as a function of the central value of the
    contrast in the energy density for the same evolution as in 
    Figures~\ref{fig:Dust_L2.jmpot} and~\ref{fig:Dust_L2.jmcont}.  
    The dashed line corresponds to
    the results of the numerical evolution, while the solid line shows
    the analytic solution obtained in the linear regime.  The dotted
    line represents the asymptotic value for the linearized
    approximation at the origin, $C_1(r=0)$.}
  \label{fig:Dust_L2.albertocentral}
\end{figure}

Finally, in Figure~\ref{fig:Dust_L2.albertocentral} we show, for the
same initial data as in Figures~\ref{fig:Dust_L2.jmpot}
and~\ref{fig:Dust_L2.jmcont}, the evolution of the central value of
the potentials $\Psi_{\rm g.i.}$ and $\Phi_{\rm g.i.}$ as a function
of the contrast in the energy density evaluated at the center of the
configuration.  Note that the two potentials still follow each other
closely during the evolution, but their values start to differ from
the linear theory prediction as time moves forward.  Moreover, while
in the linear approximation the central absolute value of the
potentials continues to become smaller, in the nonlinear evolution
this absolute value first becomes smaller, but later starts to grow
again, the transition occurring roughly when the maximum of the energy
contrast has reached
$(\delta\rho_{\textrm{g.i.}}/\bar{\rho})_{\textrm{max}}\sim 0.6$.  It
is not surprising to find a different behavior at this level, since
the gauge-invariant quantities are only well defined during the linear
regime, and we should be cautious on how to extract meaningful results
once the nonlinear regime has been reached.  However, it is
interesting to note that top-hat models of structure formation suggest
that this transition should happen at a density contrast of order
$(\delta\rho_{\textrm{g.i.}}/\bar{\rho})_{\textrm{max}} \sim 1.7$,
which corresponds to the moment when the core starts to collapse as a
closed Friedmann universe (see for instance Section 9.3 in
Ref.~\cite{Lyth:2009zz}). Our results show that, at least for the initial
conditions considered here, this transition occurs well below this
value.

\section{Real massive scalar field}
\label{sec.scalar.field}

In this section we consider the evolution of a universe dominated by a
massive scalar field satisfying the Klein-Gordon equation,
\begin{equation}
\left( \Box - m^2 \right) \varphi = 0 \; .
\end{equation}
In the above equation $m$ is the ``mass parameter'' of the scalar
particle and has dimensions of inverse distance (it is just the
inverse of the Compton wavelength of the scalar
particle).~\footnote{If the fundamental particle associated with the
  scalar field has mass $\mu$, then we have $m=\mu c/ \hbar = 1
  /\lambda_{\rm C}$, with $\lambda_{\rm C}$ the Compton wavelength, and where for
  the sake of clarity we have written explicitly the dependence on the
  speed of light $c$. The mass parameter therefore provides a natural
  length-scale for the scalar field.} The energy-momentum tensor
associated to a canonical, real, massive scalar field with no
self-interactions takes the form
\begin{equation}
T_{\mu\nu} = \partial_{\mu}\varphi\partial_{\nu}\varphi
- \frac{1}{2} g_{\mu\nu} \left( \partial_{\alpha} \varphi \partial^{\,\alpha} \varphi
+ m^2 \varphi^2 \right) \, .
\end{equation}

A free scalar field oscillating with high frequency (when compared to
the expansion rate of the universe) around the minimum of the
potential has been considered as a possible candidate to describe the
dark matter content of the universe~\cite{Dine:1982ah, Matos:2000ss,
  Marsh:2010wq}.  To leading order in $H/m$, the energy density of the
coherent oscillations decays with the cosmological expansion as
$\rho\sim 1/a^3$, with $a\sim \eta^2$~\cite{Dine:1982ah, Turner:1983he, Alcubierrepreparation}.
The linear order perturbations evolve like those of a pressureless
perfect fluid, except for the fact that the dynamics of the field
introduce a cutoff in the power spectrum of the density perturbations
at the scale of the Compton wavelength of the scalar particle,
$\lambda_{\rm C}\sim 1/m$, so that modes with smaller wavelength do not grow
but just disperse with decaying amplitude, as we will show
below~\cite{Alcubierrepreparation}.

In order to test this picture we evolve numerically cosmological
initial data with a scalar field as the matter content, using the same
full nonlinear code that we employed in Section~\ref{sec.perfect
  fluid}. For small perturbations we corroborate the analytical
results obtained using the linearized theory. Furthermore, when the
perturbations reach the nonlinear regime we find that, in general,
they grow at a slower rate than in the case of dust. This is expected
to delay the formation of the first structures in the universe.

To deal with the numerical evolution of the scalar field we reduce the
Klein-Gordon equation to a first-order form by introducing the
auxiliary variables $\Pi:=n^{\mu}\partial_{\mu}\varphi$ and
$\Theta_i:=\partial_i \varphi$. The Klein-Gordon equation can then be
written as the following system of evolution equations
\begin{subequations}
\begin{eqnarray}
\label{eq:ev_sf_phi}
\p_t \varphi\; - \Lie{\beta}{\,\varphi\;} &=&  \alpha \Pi \; , \\
\label{eq:ev_sf_Psii}
\p_t \Theta_i - \Lie{\beta}{\Theta}_i &=& \p_i (\alpha \Pi) \; , \\
\label{eq:ev_sf_pi}
\p_t \Pi\, - \Lie{\beta}{\,\Pi\,} &=& \alpha \Pi K 
- \alpha m^2 \varphi + \nabla_i\left(\alpha\Theta^i \right) \,.
\end{eqnarray}
\end{subequations}
These equations are solved simultaneously with the integration of the
spacetime system. In terms of the new variables the sources of the
gravitational field given in Eqs.~(\ref{source.grav.1})
and~(\ref{source.grav.2}) take the form
\begin{subequations}\label{initial.scalar1}
\begin{eqnarray}
\rho &=& \frac{1}{2} \left(\Pi^2 + \Theta_m\Theta^m + m^2 \varphi^2 \right)\,,\\
j_i &=& -\Pi\Theta_i \,, \label{eq.momentum.scalar}\\
S_{ij} &=& \Theta_i \Theta_j + \frac{1}{2} \left(\Pi^2 - \Theta_m\Theta^m
- m^2\varphi^2 \right) \gamma_{ij} \,.
\end{eqnarray}
\end{subequations}
For generality, the above expressions have been written for the
three-dimensional case.  In our code they are reduced to their
spherically symmetric form, where $\Theta_i$ only has a radial
component.

As we did for the case of a perfect fluid, we perform numerical
evolutions with a vanishing shift vector, and a slicing condition of
the Bona-Masso family with \mbox{$f=0.3332$}.  Again, we consider
initial data with vanishing momentum density by
choosing $\Pi(t=t_0)=0$.
With these assumptions the initial data in Eqs.~(\ref{eq.initial
  data.2}) simplify to (remember that in
Section~\ref{sec:num.evolutions.II} we already fixed
$\hat{\gamma}_{\|0}=1$)
\begin{subequations}
\begin{eqnarray}
\rho(t=t_0,r) &=& \frac{1}{2} \left[ (\psi^{-2}_0 \Theta_{r0})^2
+ m^2 \varphi^2_0 \right] \,, \\
j_r (t=t_0,r) &=& 0 \,, \\
S_{\|}(t=t_0,r) &=& \frac{1}{2} \left[ (\psi^{-2}_0\Theta_{r0})^2
- m^2 \varphi^2_0 \right] \,, \\
 S_{\perp}(t=t_0,r) &=& \frac{1}{2} \left[ -(\psi^{-2}_0\Theta_{r0})^2
- m^2 \varphi^2_0 \right] \,.
\end{eqnarray}
\end{subequations}
Note that for the homogeneous and isotropic configurations,
$\Theta_{r0}=0$, we recover $S_{\|0}=S_{\perp0}$, as expected from
Eqs.~(\ref{eq.initial data.1}) above.

Contrary to the case of dust in Section~\ref{sec.perfect fluid}, we
now do not specify the initial profile for the energy density,
$\rho_0(r)$, but rather the initial scalar field profile $\varphi_0$.
In particular, we consider initial data of the form
\begin{subequations}\label{eq.scalar.initial}
\begin{equation}
 \varphi_0(r) = \bar{\varphi}_0 + \delta\varphi_0(r) \,,
\end{equation}
with 
\begin{equation}\label{par.pertur2sf}
\delta\varphi_0(r) = \delta\varphi_{*} \left(1 - \frac{11 r^2}{3L^2} \right)
\left[ \left( 1 - \frac{r}{L} \right) \left( 1 + \frac{r}{L} \right) \right]^3
\end{equation}
\end{subequations}
for \mbox{$0<r<L$}, and $\delta\varphi_0(r)=0$ otherwise. Again, the
parameter $L$ represents a length-scale for the perturbation, whereas
$\delta\varphi_{*}$ measures the amplitude of the inhomogeneity in the
scalar field, both evaluated at $t=t_0$. Notice that since the leading-order 
contribution to the energy density perturbation will be linear
in $\delta\varphi_0$, the above choice still guarantees that to first
order the density perturbation will integrate to zero (although
we will have nonvanishing second-order contributions to the
integral).  A geometry consistent with this initial distribution of
matter is obtained following the recipes of
Sections~\ref{sec:num.evolutions} and~\ref{sec:num.evolutions.II}.

\begin{figure}[t]
  \centering
  \includegraphics[width=0.49\textwidth]{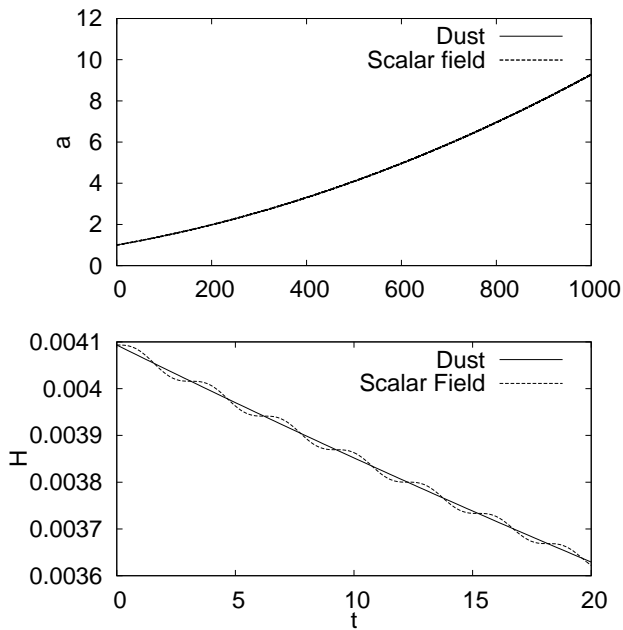}
  \caption{Evolution of the scale factor $a$ (top panel) and Hubble
    parameter $H$ (bottom panel), as functions of coordinate time $t$, for an expanding
    homogeneous and isotropic universe dominated by the coherent
    oscillations of a massive scalar field (dotted line).  Here we
    have chosen $\varphi_0= 2\times 10^{-3}$ and $\delta\varphi_*=0$ in order
    to have the same initial configuration for the scale factor as
    in Figure~\ref{fig:Dust_a_evol}. 
    We also show the analytic expression for the case of dust,
    Eq.~(\ref{eq:scale.factor.dust}) (solid line). The scale 
    is zoomed in near the origin for the bottom panel in order to
    appreciate the oscillations of the universe containing the
    scalar field.}
  \label{fig:SF_a_evol}
\end{figure}

Figure~\ref{fig:SF_a_evol} shows, for a scalar field of mass $m=1$ and
initial data $\bar{\varphi}_0=2\times10^{-3}$, $\delta\varphi_{*}=0$,
a comparison of the evolution of the scale factor obtained with our
numerical implementation against the analytical solution for a
Friedmann universe dominated by dust,
Eq.~(\ref{eq:scale.factor.dust}). In this case the scalar field has
already settled to oscillate near the minimum of potential, and drives
the evolution of the scale factor in a way that mimics very well the
behavior obtained in the case of dust, Figure~\ref{fig:Dust_a_evol} above.
Nevertheless, on short time-scales there are small oscillations (of order
$H/m$) of the dynamical quantities around the dust solution. These oscillations are
better appreciated on the Hubble factor (see the lower panel in Figure~\ref{fig:SF_a_evol}). 
Apart from the small oscillations the two solutions coincide up to discretization
error.

\begin{figure}[t]
  \centering
  \includegraphics[width=0.49\textwidth]{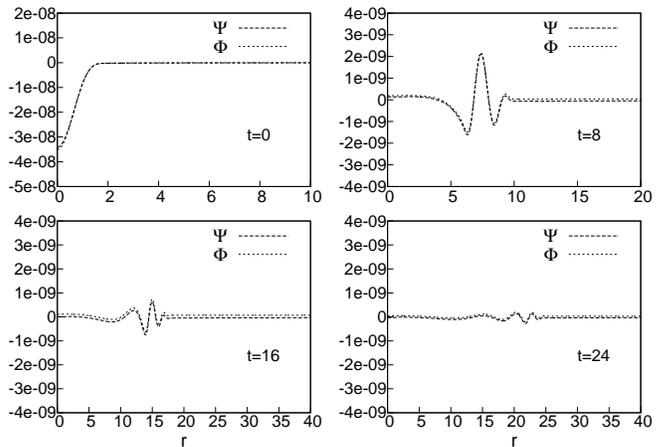}
  \caption{Snapshots of the gauge-invariant perturbations in the
    metric tensor, $\Phi_{\textrm{g.i.}}$ and
    $\Psi_{\textrm{g.i.}}$, as a function of the comoving radial
    coordinate, $r$, at different instants of time in the cosmological
    evolution.  Here we have chosen an initial data of the form in
    Eq.~(\ref{eq.scalar.initial}) with $\bar{\varphi}_0=2.0\times
    10^{-3}$, $\delta\varphi_*=10^{-5}$, and \mbox{$L=2$}.  Note that
    the inhomogeneity is composed mostly of short wavelength modes
    that disperse away.}
  \label{fig:SF_L.small}
\end{figure}

\begin{figure}[t]
  \centering
  \includegraphics[width=0.49\textwidth]{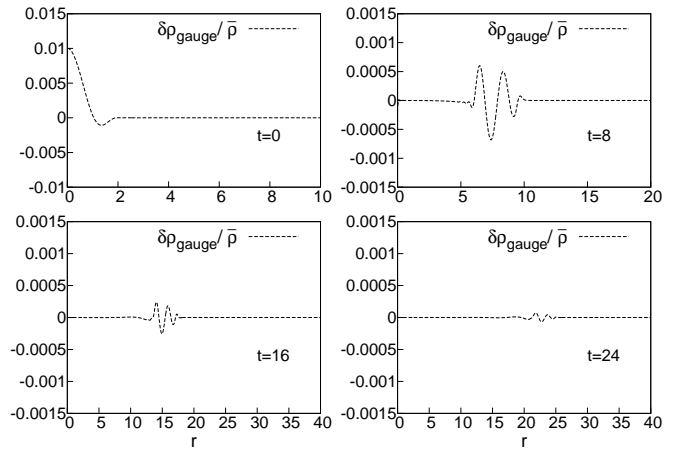}
  \caption{Snapshots of the energy density contrast,
    $\delta\rho_{\textrm{g.i.}}/\bar{\rho}$, for the same initial
    data as in Figure~\ref{fig:SF_L.small}. 
    }
  \label{fig:SF_L.small2}
\end{figure}

To analyze the behavior of the small perturbations we need to consider
two different scales relative to the Compton wavelength of the scalar
particle. 
Short-scaled perturbations (whose characteristic lengths are of order
$L\lesssim 1/m$) are composed mostly of
modes shorter than the Compton wavelength. These modes propagate
freely like (almost) a massless scalar field, and these perturbations
are not expected to collapse unless their initial amplitude is very
large.  As an example of this behavior, in
Figures~\ref{fig:SF_L.small} and~\ref{fig:SF_L.small2} we show the
numerical evolution of the gauge-invariant perturbations in the metric
tensor, $\Phi_{\textrm{g.i.}}$ and $\Psi_{\textrm{g.i.}}$, and the
energy density contrast, $\delta\rho_{\textrm{g.i.}}/\bar{\rho}$, for
a scalar field with an initial data characterized by the parameters
$\bar{\varphi}_0=2\times 10^{-3}$, $\delta\varphi_{*}=10^{-5}$, and
$L=2$.
We see that basically the whole initial perturbation disperses leaving
at $t=15$ only a tiny imprint on the physical quantities measuring
the inhomogeneities.  

\begin{figure}[t]
  \centering
  \includegraphics[width=.49\textwidth]{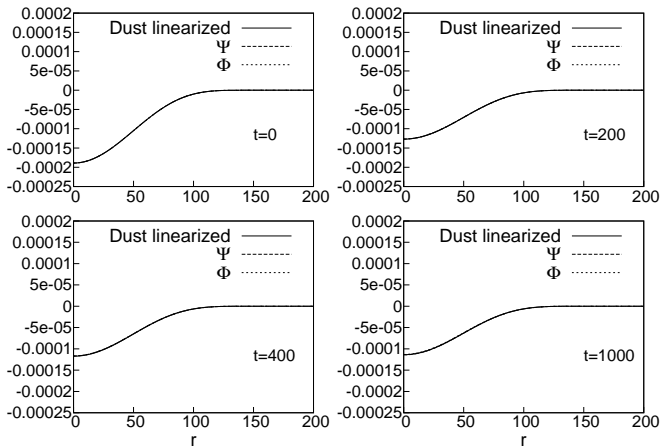}
  \caption{Similar to Figure~\ref{fig:SF_L.small} but now for an
    initial data with $L=150$.  We show the results of the numerical
    evolution of the gauge-invariant quantities
    $\Psi_{\textrm{g.i.}}$ (dotted-line) and $\Phi_{\textrm{g.i.}}$
    (dashed-line), together with the analytic solution obtained in the
    linear regime for a universe dominated by dust,
    Eq.~(\ref{eq.Psi.linear}) (solid-line).}
  \label{fig:SF_L.intermediate}
\end{figure}

\begin{figure}[t]
  \centering
  \includegraphics[width=.49\textwidth]{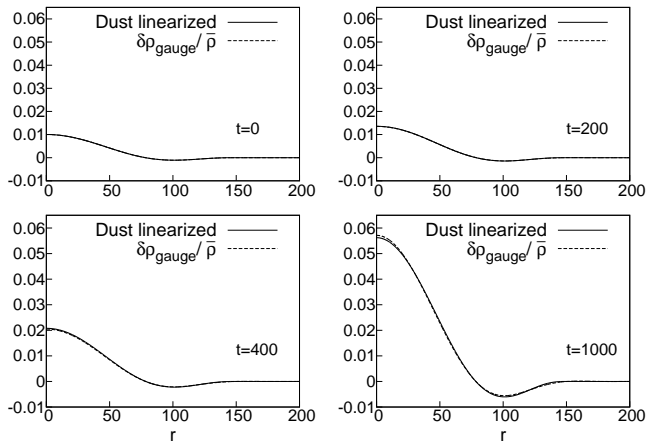}
  \caption{Similar to Figure~\ref{fig:SF_L.small2} but now for an
    initial data with $L=150$.  We show the results of the numerical
    evolution of the density contrast (dashed-line), together with
    the analytic solution obtained in the linear regime for a universe
    dominated by dust, Eq.~(\ref{eq.contrast.linear}) (solid-line).}
  \label{fig:SF_L.large}
\end{figure}

On the other hand, large-scaled perturbations with $L \gg 1/m$ have a
significant contribution from modes larger than the Compton
wavelength.  These modes grow in a very similar way to those in a
universe dominated by dust~\cite{Alcubierrepreparation}. In
Figures~\ref{fig:SF_L.intermediate} and~\ref{fig:SF_L.large} we show,
for initial data with the same values of $\bar{\varphi}_0$ and
$\delta\varphi_{*}$ as in Figures~\ref{fig:SF_L.small}
and~\ref{fig:SF_L.small2} but with $L=150$, the evolution of the gauge
invariant perturbations of the metric and the energy density
contrast, respectively, together with the predictions of the
linearized theory for the case of dust.  Note that in both figures the
numerical evolution for the case of the scalar field coincides with
that expected for dust.

\vspace{5mm}

\begin{figure}[ht]
  \centering
  \includegraphics[width=0.49\textwidth]{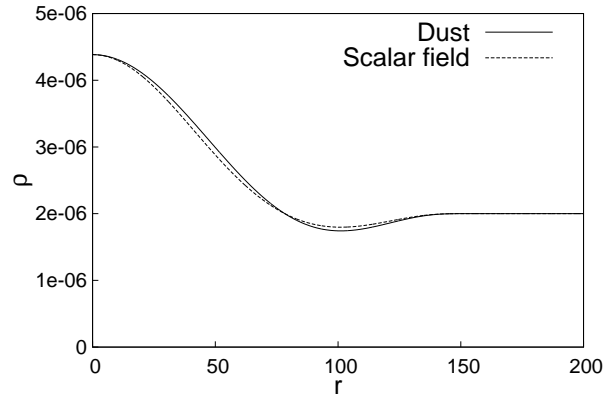}
  \caption{Initial density profiles used for the evolutions starting
    in the full nonlinear regime, Figures~\ref{fig:race1} and~\ref{fig:race2} 
    below. The parameters were adjusted in
    order to have initially the same central value of the energy density in dust (solid-line)
    and in the scalar field (dashed-line).}
\label{fig:12b_rho.comparison}
\end{figure}

Let us conclude this paper with a comparison of the nonlinear behavior
of the perturbations, for which analytical results do not exist, for
the cases of a universe dominated by dust and one dominated by a
scalar field. In order to start from a situation that is clearly
outside the linear regime, we consider initial data such that the
value of the energy density at the origin is roughly twice the value
of the energy density in the asymptotic homogeneous region. For the
case of a universe driven by dust we set the parameters
$\bar{\rho}_0=2.0\times 10^{-6}$ and $\delta\rho_{*}=2.38\times
10^{-6}$ in Eq.~(\ref{eq.density.initial.gen}), while for the scalar
field we take $\bar{\varphi}_0=2\times 10^{-3}$ and
$\varphi_{*}=9.611\times10^{-4}$ in Eq.~(\ref{eq.scalar.initial}). In
both settings the inhomogeneity has a length-scale of $L=150$. One
should notice that, since we are starting in the nonlinear regime, the
construction of the initial data for the scalar field does not result
in the same energy density profile as in the case of dust. However,
the configurations constructed with these values are in fact very
similar and good enough for the comparison (see
Figure~\ref{fig:12b_rho.comparison} for details).

The results we show in this case do not correspond directly with the
ones observed in the linear regime, since some of the quantities used
in that case are now not well defined. This occurs since once we are
outside the linear regime the dependence on the choice of slicing
becomes relevant. The asymptotic region still evolves as a true
background, so long as it is casually disconnected from the
inhomogeneities in the central regions, and then quantities measured by
distant observers at rest with respect to this asymptotic background
are common to all of them. The question that arises is how we can
compare the properties of the spacetime in the central regions to
those of the background. It turns out that in spherical symmetry one
can consider for this purpose the physical quantities measured by an
observer at the origin as functions of proper time, since the
spacetime trajectory of this observer is gauge independent for
symmetry reasons.

\begin{figure}[t]
  \begin {center}
   \includegraphics[width=.49\textwidth]{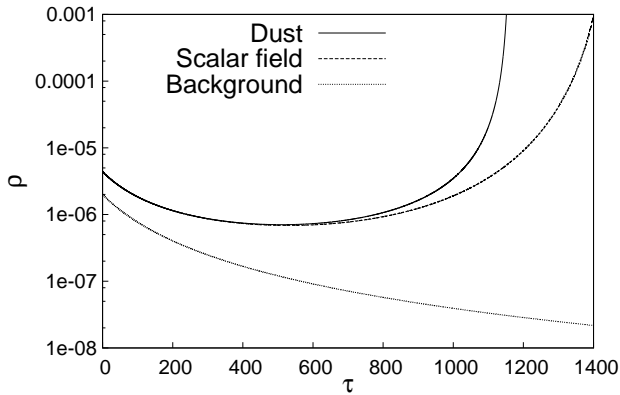}
   \caption{Evolution of the central energy density measured by the
     Eulerian observers as a function of proper time $\tau$, for both
     a universe dominated by dust (solid-line) and one dominated by a scalar
     field (dashed-line). For comparison we also show the value of the energy
     density on the boundary of the computational domain (dotted-line). In both
     cases the central energy density deviates from the asymptotic
     background evolution, reaching a turnaround point and finally
     undergoing full gravitational collapse. Meanwhile, the
     asymptotic region evolves as a true background through the whole
     simulation.}
    \label{fig:race1}
  \end {center}
\end{figure}

Based on these observations, to study how the inhomogeneity collapses
we show in Figure~\ref{fig:race1} the evolution of the energy density
measured by the Eulerian observers as a function of their proper time,
evaluated both at the origin and at the boundary of the computational
domain.  Initially, the energy density at the origin becomes smaller
at a similar rate for both the scalar field and dust, as a consequence
of the expansion of the universe. However, after some time the central
energy density reaches a turnaround point. Both configurations
eventually undergo gravitational collapse, forming (small) apparent
horizons.  Soon thereafter our simulations fail both because our
slicing condition is not well adapted to the evolution of black hole
spacetimes, and also because our resolution in the central regions is
not high enough. In the case of dust the collapse proceeds directly,
but for the scalar field once the central density becomes large the
effect of the local oscillations dominates so that the collapse to a
black hole is delayed.~\footnote{Although the details of the collapse
  to a black hole are not the aim of this paper, it would nevertheless
  be interesting to study if for the case of a scalar field there is
  some transient metastable configuration before a horizon forms.
  For this one would need to use a better slicing condition that can
  work well both for the asymptotic background and for the central
  collapsing regions.}

\begin{figure}[ht]
  \begin {center}
   \includegraphics[width=.49\textwidth]{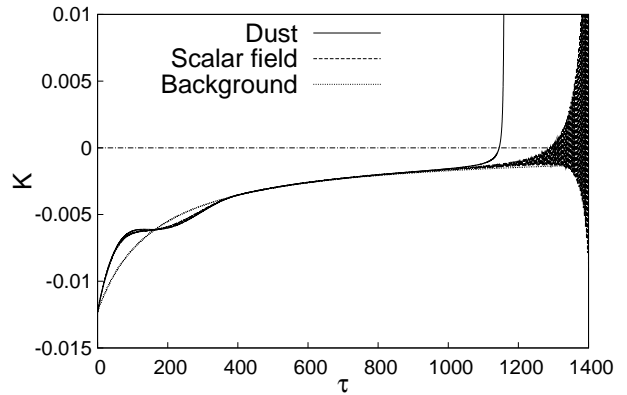}
   \caption{Central value of the trace of the extrinsic curvature $K$
     (essentially the negative of the expansion) as a function of
     proper time for both dust (solid-line) and scalar field (dashed-line) evolutions. For
     comparison, the value on the background asymptotic region is also
     plotted (dotted-line). After a transient when local effects dominate, the value
     of the expansion catches on with that of the background and
     deviates again dramatically when reaching a state of full
     gravitational collapse.}
    \label{fig:race2}
  \end {center}
\end{figure}

We also show in Figure~\ref{fig:race2} the evolution of the trace of
the extrinsic curvature at the origin, which encodes the local
expansion of the volume elements (one is proportional to the negative
of the other). For comparison the value at the boundary is also
plotted, all as a function of proper time. Our prescription of initial
data assumes that the expansion is initially identical in all the
domain, but as soon as the evolution starts the central expansion
decreases faster in the over-dense regions.  After some transient,
however, the effect of our slicing condition causes the expansion at
the center to follow closely that of the background. Finally, once the
central regions are in a state of full gravitational collapse the
expansion deviates again from the background and changes sign. As
already discusses above, for the case of dust this leads directly to a
black hole, but in the case of the scalar field the effect of the
local oscillations become dominant as can be clearly seen in the
figure.

\section{Conclusions}
\label{sec:Discussion}

We have used a numerical relativity code in order to follow the
evolution of cosmological scenarios with inhomogeneous, spherically
symmetric distributions of matter. When the inhomogeneities are small,
both the background energy density and its perturbation decrease as
the universe expands, and only the density contrast can grow. In our
simulations we have been able to reach and surpass the so-called
turn-around point, where the nonlinear terms dominate and the
inhomogeneities themselves start to collapse.

We have considered two different situations: a universe dominated by a
perfect fluid with no pressure, {\em i.e.} dust, and one dominated by the
coherent oscillations of a massive real scalar field around the
minimum of the potential. In both these scenarios matter is described
in terms of a field theory and N-body simulations are not viable.

For the case of dust there is a well-known analytic solution that
describes the behavior of the homogeneous and isotropic background
as well as its small perturbations,
and we have shown that our numerical evolutions can reproduce such a
solution with high accuracy.  For the case of a universe dominated by a
scalar field there are no exact analytic expressions. However at late
times, when the mass of the scalar particle is much larger than the
Hubble parameter, $m \gg H$, our numerical simulations show that the
scale factor just oscillates around the solution corresponding to
dust, as was already expected from previous approximate
studies~\cite{Dine:1982ah}.  These oscillations have small amplitude
and high frequency, and damp with the cosmological expansion.

The linear perturbations in the case of the scalar field also evolve
in a very similar way as those for dust, as long as their initial size
is larger than the Compton wavelength of the scalar particle,
$\lambda_{\rm C}=1/m$.  On the other hand, initial perturbations in the
scalar field with a length-scale of the order of $\lambda_{\rm C}$ or shorter do
not grow and instead propagate freely with decaying amplitude. This
introduces a cutoff in the power spectrum of the scalar field, as has
been previously pointed out by other authors in
{\em e.g.} Refs.~\cite{Ratra:1988bz, Hwang:1996xd, Matos:2000ng, Matos:2000ss,
  Marsh:2010wq}.  Once in the nonlinear regime, in the case of dust
the initial inhomogeneity simply continues to collapse under its own
gravity until a black hole is formed.  For the case of the scalar
field this final collapse proceeds in a different way: once the
central density becomes large the strong local oscillations in the
field dominate and the collapse to a black hole is delayed.

The results presented here support the idea that when considering the
coherent excitation of a scalar field as a dark matter candidate it is
possible to recover similar cosmological large scale structure
formation to that obtained with the standard cold dark matter
model. Some differences are observed, namely those related to the
finite cutoff at short scales, the small high frequency oscillations
in the scale factor, and the final state of the gravitational collapse,
that may have observational consequences.

Finally we mention some ideas that we leave for future works. First
among them there is the problem of forming nonsingular, stable
objects starting from inhomogeneous distributions of matter in
expanding universes. For the case of dust there is no room for an
outcome different to collapse to a black hole. However, if matter is
described by the coherent excitation of a scalar field there are well
known regular and stable self-gravitating configurations: oscillatons
if the scalar field is real~\cite{Seidel:1991zh,Alcubierre03a}, and
boson stars for the case of a complex
field~\cite{Ruffini69,Liebling:2012fv}.  It is an open question
whether the gravitational collapse of a scalar field in a cosmological
context can result in such stable configurations for some choices of
the initial data parameters, although this is what one would expect in 
a successful model of dark matter.  On a more practical level, it would be
also important to explore different slicing conditions that might
allow for longer evolutions in the collapsing regions, as well as to
extend our numerical code in order to describe new types of matter (a
collection of noncollisional particles described by the Vlasov
equation, for example) and more general perturbations in full 3D.  
In any case, we believe that the study of cosmological
nonlinear structure formation within the realm of full general
relativity can shed new light on the nature of the mysterious dark
components of the universe.


\acknowledgments

We are grateful to Tonatiuh Matos, Olivier Sarbach and Luis Ure\~na-Lopez for useful comments on a first draft of this paper.
This work was partially supported by DGAPA-UNAM under Grants No.~IN115311
and No.~IN103514, and CONACyT Mexico No.~182445 and No.~167335.  JMT acknowledges
CONACyT for a graduate grant. 
ADT is supported in part by Grant No.~FQXi-1301 from the Foundational Questions Institute (FQXi).


\appendix


\bibliography{bibtex/referencias}
\bibliographystyle{bibtex/prsty}


\end{document}